\documentclass[aip,reprint,showpacs,superscriptaddress,groupedaddress,balancelastpage,floatfix]{revtex4-2}  

\usepackage[utf8]{inputenc}
\usepackage{graphicx,dcolumn,bm,amssymb,amsmath,amsfonts,xcolor}
\usepackage[
pdfstartview=FitBV,
bookmarks=true,
bookmarksopen=true,
colorlinks =true,
linkbordercolor=blue,
linkcolor = blue,
citecolor = blue,
urlcolor=blue]{hyperref}
\usepackage[capitalize]{cleveref} 

\crefname{section}{Sec.}{Sections}
\crefname{table}{TABLE}{TABLEs.}
\crefname{figure}{FIG.}{FIGs.}

{}
\begin{document}
	
	\title{Classical threshold law for the formation of van der Waals molecules}
	\author{Marjan Mirahmadi}
	\email{mirahmadi@fhi-berlin.mpg.de}
	\affiliation{Fritz-Haber-Institut der Max-Planck-Gesellschaft, Faradayweg 4-6, D-14195 Berlin, Germany}
	\author{Jes\'us P\'{e}rez-R\'{i}os}
	\email{jperezri@fhi-berlin.mpg.de}
	\affiliation{Fritz-Haber-Institut der Max-Planck-Gesellschaft, Faradayweg 4-6, D-14195 Berlin, Germany}
	\date{\today}
	
		\begin{abstract}
		We study the role of pairwise long-range interactions in the formation of van der Waals molecules through direct three-body recombination processes A + B + B $\rightarrow$ AB + B, based on a classical trajectory method in hyperspherical coordinates developed in our earlier works [J. Chem. Phys. \textbf{140}, 044307 (2014); J. Chem. Phys. \textbf{154}, 034305 (2021)]. In particular, we find the effective long-range potential in hyperspherical coordinates with an exact expression in terms of dispersion coefficients of pairwise potentials. Exploiting this relation, we derive a classical threshold law for the total cross section and the three-body recombination rate yielding an analytical expression for the three-body recombination rate as a function of the pairwise long-range coefficients of the involved partners. 
		
	\end{abstract}
	
	\maketitle
	
	\section{Introduction}\label{sec:intro}
	Van der Waals (vdW) molecules consist of two atoms held together by the long-range dispersion interaction~\cite{Blaney1976} leading to (ground state) binding energies $\lesssim$1~meV. Thus, setting aside ultra-long-range Rydberg molecules (with binding energies $\sim 4$~neV~ \cite{Greene2000,Hamilton2002,Bendkowsky2009,Booth2015,Niederprum2016}), vdW molecules show the weakest gas-phase molecular bond in nature. The binding mechanism in vdW molecules is the result of the compensation between the short-range repulsion, due to the overlap of closed-shell orbitals, and the attractive vdW interaction ($-C_6/r^6$) caused by zero point fluctuations of atomic dipole moments.
	
	The study of vdW interactions provides a deeper understanding of crucial phenomena in physics, chemistry, and biology. For instance, these interactions play a key role in the formation and stability of gases, liquids, vdW heterostructures and biopolymers;~\cite{Buckingham1988,Koperski2002,Hermann2017} chemical reactions;~\cite{Smalley1977,Worsnop1986,Balakrishnan2004,Shen2017} superfluidity of $^{4}$He nanodroplets; ~\cite{Toennies2004,Szalewicz2008} and rare gas crystals and the dynamics of impurities interacting with dense rare gas vapors.~\cite{Fugol1978,Brahms2011}  
	
	Recent developments in cooling techniques, specifically buffer gas cooling,~\cite{DeCarvalho1999} have paved the way to new possibilities for investigating the formation of vdW molecules through three-body recombination processes.\cite{Brahms2008,Suno2009,Brahms2010,Brahms2011,Wang2011,Tariq2013,Quiros2017} Three-body recombination is a three-body collision during which two of the particles form a bound state. These reactions play an important role in a wide range of physical and chemical phenomena, ranging from H$_2$ formation in star-forming regions~\cite{Flower2007,Forrey2013} to loss mechanisms in ultracold dilute atomic gases~\cite{Esry1999,Weiner1999,Bedaque2000,Suno2003,Weber2003,Schmidt2020,Greene2017} to the formation and trapping of cold and ultracold molecules.~\cite{Koehler2006,Blume2012,Perez-Rios2015,Krukow2016,Mohammadi2021}
	
	In Ref.~[\onlinecite{Mirahmadi2021}], we considered the formation of atom-rare gas vdW molecules via a direct three-body recombination mechanism at temperatures relevant for buffer gas cell experiments. As a result, we found that almost any atom in a helium buffer gas will evolve into a vdW molecule. Fueled by those results, our goal in the present work is to present a comprehensive study of A + B + B reactions and derive a classical threshold law for the formation of vdW molecules in cold environments. To investigate this problem, we use a classical approach in hyperspherical coordinates, which has previously been used to consider the three-body recombination of three neutral atoms,~\cite{Perez-Rios2014,Greene2017,Mirahmadi2021} as well as ion-neutral-neutral three-body recombination processes.~\cite{Perez-Rios2015,Krukow2016,Perez-Rios2018} In order to derive a threshold law, we have obtained an effective long-range potential in hyperspherical coordinates. With it, a general expression (as a function of $C^\mathrm{AB}_6$ and  $C^\mathrm{B_2}_6$) for the corresponding dispersion coefficient is given by considering several A and B atoms chosen from alkali metals, alkaline-earth metals, transition metals, pnictogens, chalcogens, halogens, and rare gases. Furthermore, we calculated the threshold values for the three-body recombination rates at 4~K. Our results confirm that any vdW molecule AB appears with almost the same probability.

	This paper is organized as follows: In \cref{sec:0}, the Hamiltonian governing the classical dynamics during three-body recombination in the three-dimensional space and its counterpart in the six-dimensional space are introduced. In \cref{sec:1}, the long-range potential in hyperspherical coordinates have been obtained, and the relevant, effective dispersion coefficient as a function of dispersion coefficients of the pairwise interactions is found. In \cref{sec:2}, a classical threshold law \`a la Langevin for the total cross section and the three-body recombination rate are developed. Finally, in \cref{sec:conclusion}, we summarize our chief results and discuss their possible applications.

	\section{Three-body recombination in hyperspherical coordinates}\label{sec:0}
\begin{figure}
		\begin{center}
			\includegraphics[scale=0.2]{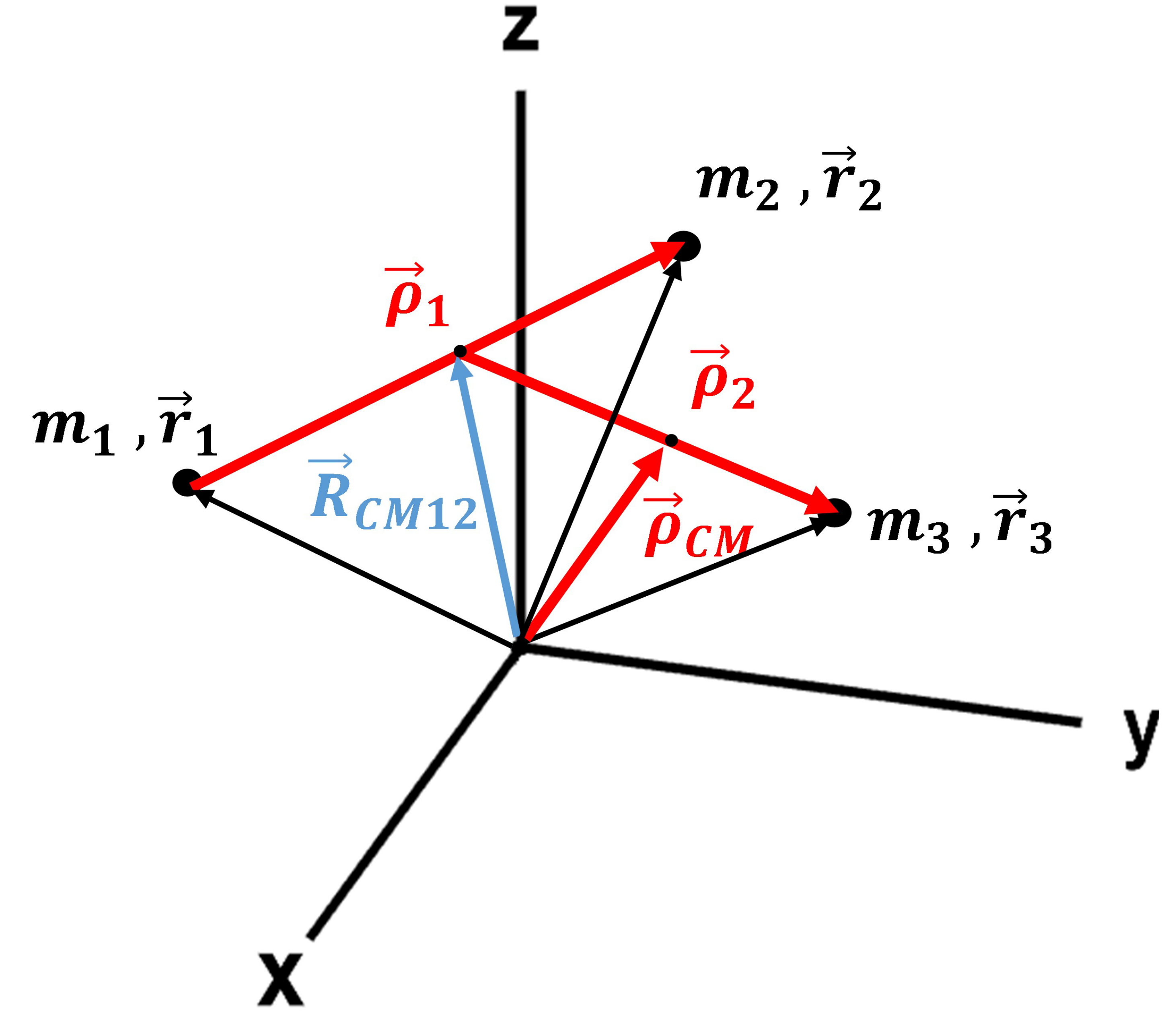}
			\caption{Jacobi coordinates for the three-body problem illustrated by the red vectors. Black arrows indicate the position of the three particles in Cartesian coordinates and the blue arrow indicates the two-body center-of-mass vector $\vec{R}_{CM12}$.}	
			\label{fig:jacobi}	
		\end{center}
	\end{figure}

	Consider a system consisting of three particles with masses $m_i$ ($i = 1,2,3$) at  positions $\vec{r}_i$, interacting with each other via the potential $	V(\vec{r}_1,\vec{r}_2,\vec{r}_3)$. The motion of these particles is governed by the Hamiltonian
	\begin{equation}\label{eq:cartesianH}
		H = \frac{\vec{p}_1^{~2}}{2m_1} + \frac{\vec{p}_2^{~2}}{2m_2} + \frac{\vec{p}_3^{~2}}{2m_3} + V(\vec{r}_1,\vec{r}_2,\vec{r}_3) ~,
	\end{equation}
	with  $\vec{p}_i$ being the momentum vector of the $i$-th particle. It is more convenient to treat the three-body problem in Jacobi coordinates~\cite{Pollard1976,suzuki1998} defined by the following relations	
	\begin{align}\label{eq:jacobitrans}
		\vec{\rho}_1 &= \vec{r}_2 - \vec{r}_1 ~, \nonumber \\
		\vec{\rho}_2 &= \vec{r}_3 -\vec{R}_{CM12} ~, \nonumber \\
		\vec{\rho}_{CM} &= \frac{m_1\vec{r}_1 + m_2\vec{r}_2 + m_3\vec{r}_3}{M} ~,
	\end{align}
	where $ \vec{R}_{CM12} = (m_1\vec{r}_1 + m_2\vec{r}_2)/(m_1+m_2)$  is the center-of-mass vector of the two-body system consisting of $m_1$ and $m_2$. $M = m_1 + m_2 + m_3$ and $\vec{\rho}_{CM}$ are the total mass and the center-of-mass vectors of the three-body system, respectively. The Jacobi vectors are illustrated in \cref{fig:jacobi}.
	
	Due to conservation of the total linear momentum (i.e., $\vec{\rho}_{CM}$ is a cyclic coordinate), we omit the degrees of freedom of the center of mass and write the Hamiltonian~\eqref{eq:cartesianH} as
	\begin{equation}\label{eq:jacobiH}
		H = \frac{\vec{P}_1^2}{2\mu_{12}} + \frac{\vec{P}_2^2}{2\mu_{3,12}} +  V(\vec{\rho}_1,\vec{\rho}_2) ~,
	\end{equation} 
	with reduced masses $\mu_{12}=m_1m_2/(m_1 + m_2)$ and $\mu_{3,12}=m_3(m_1+m_2)/M$. Here,  $\vec{P}_1$ and $\vec{P}_2$ indicate the conjugated momenta of $\vec{\rho}_1$ and $\vec{\rho}_2$, respectively. Note that, since the relations given by \cref{eq:jacobitrans} indicate a canonical transformation, the Hamilton's equations of motion are invariant under the transformation to Jacobi coordinates. 
	
	\subsection{Hyperspherical coordinates}\label{subsec:HSC}
	In the next step, we map the independent relative coordinates of the three-body system associated with the Hamiltonian~\eqref{eq:jacobiH} in a three-dimensional (3D) space, onto the degrees of freedom of a single particle moving towards a scattering center in a six-dimensional (6D) space under the effect of the Hamiltonian $H^\mathrm{6D}$. This 6D space is described by means of hyperspherical coordinates consisting of a hyperradius $R$, and five hyperangles $\alpha_j$ (with $j = 1,2,3,4,5$), where $0\leq\alpha_1<2\pi$ and $0\leq\alpha_{j>1}\leq\pi$.~\cite{Lin1995,Avery2012} The components of a 6D vector $\vec{x}=(x_1,x_2,x_3,x_4,x_5,x_6)$ in hyperspherical coordinates are given by
	\begin{align}\label{eq:hsCart}
		x_1 & = R \sin(\alpha_1)\sin(\alpha_2)\sin(\alpha_3)\sin(\alpha_4)\sin(\alpha_5) ~, \nonumber \\
		x_2 & = R  \cos(\alpha_1)\sin(\alpha_2)\sin(\alpha_3)\sin(\alpha_4)\sin(\alpha_5) ~, \nonumber \\
		x_3 & = R  \cos(\alpha_2)\sin(\alpha_3)\sin(\alpha_4)\sin(\alpha_5) ~, \nonumber \\
		x_4 & = R  \cos(\alpha_3)\sin(\alpha_4)\sin(\alpha_5) ~, \nonumber \\
		x_5 & = R  \cos(\alpha_4)\sin(\alpha_5) ~, \nonumber \\
		x_6 & = R  \cos(\alpha_5) ~,
	\end{align} 
	and the volume element in this coordinate system is given by
	\begin{align}\label{eq:hsvol}
		d\tau &= R^5d Rd\Omega \nonumber \\
		&= R^5d R\prod_{j=1}^{5}\sin^{j-1}(\alpha_j)d\alpha_j ~.
	\end{align} 
	
	The 6D position and momentum vectors can be constructed from the Jacobi vectors and their conjugated momenta as~\cite{Perez-Rios2014,Perez-Rios2020}
	\begin{equation}\label{eq:rho6D}
		\vec{\rho} = \begin{pmatrix} \vec{\rho}_1 \\ \vec{\rho}_2 \end{pmatrix} ~,
	\end{equation}
	and 
	\begin{equation}\label{eq:P6D}
		\vec{P} = \begin{pmatrix} \sqrt{\frac{\mu}{\mu_{12}}}\vec{P}_1 \\ \sqrt{\frac{\mu}{\mu_{3,12}}}\vec{P}_2 \end{pmatrix} ~,
	\end{equation}
	respectively. Here $\mu = \sqrt{m_1 m_2 m_3/ M}$ is the three-body reduced mass. Consequently,  the Hamiltonian in the 6D space reads as
	\begin{equation}\label{eq:6DH}
		H^\mathrm{6D} = \dfrac{\vec{P}^2}{2\mu} + V(\vec{\rho}) ~.
	\end{equation}
	
	\subsection{Total cross section and three-body recombination rate}\label{subsec:sigma}
	Classically, for scattering in a 3D space, the cross section $\sigma$ is defined as the area drawn in a plane perpendicular to particle's initial momentum, which the particle's trajectory should cross in order to be scattered (i.e., deviation from the uniform rectilinear motion). This concept can be extended to the 6D space by visualizing it as an area in a five-dimensional hyperplane (embedded in the 6D space) perpendicular to the initial momentum vector $\vec{P}_0$. Similarly, we define the impact parameter vector $\vec{b}$ as the projection of the initial position vector $\vec{\rho}_0$ on this hyperplane, thus the necessary condition $\vec{b}\cdot\vec{P}_0 = 0$ is satisfied.~\cite{Perez-Rios2020}
	
	Note that by treating three-body collision as a scattering problem of a single particle in a 6D space, we can uniquely define the initial conditions and the impact parameter. Hence, it is possible to obtain the probability of a three-body recombination event as a function of the impact parameter $\vec{b}$ and the initial momentum $\vec{P}_0$. Consequently, by averaging over different orientations of $\vec{P}_0$ and making use of its relation with the collision energy, $E_c=P_0^2/(2\mu)$, the total cross section of the three-body recombination process will be given by
	\begin{align}\label{eq:sigma}
		\sigma_\mathrm{rec}(E_c) & = \int \mathcal{P}(E_c,\vec{b})  b^4 d b ~d\Omega_b \nonumber \\
		&= \frac{8\pi^2}{3}\int_{0}^{b_\mathrm{max}(E_c)} \mathcal{P}(E_c,b)  b^4 d b ~,
	\end{align}
	with $d\Omega_b$ being the differential element of the solid hyperangle associated with vector $\vec{b}$. To obtain the second equality we made use of $\Omega_b = 8\pi^2/3$.  
	The function $\mathcal{P}$ in \cref{eq:sigma} is the so-called opacity function, i.e., the probability of a recombination event as a function of the impact parameter and collision energy. Note that $b_\mathrm{max}$ represents the largest impact parameter for which three-body recombination occurs, or in other words, $\mathcal{P}(E_c,b) = 0$ for $b>b_\mathrm{max}$.
	
	Finally, the energy-dependent three-body recombination rate can be achieved via the following relation
	\begin{equation}\label{eq:k3}
		k_3(E_c) = \sqrt{\frac{2E_c}{\mu}}\sigma_\mathrm{rec}(E_c) ~.
	\end{equation}
	
	\begin{figure*}
		\begin{center}
			\includegraphics[scale=0.4]{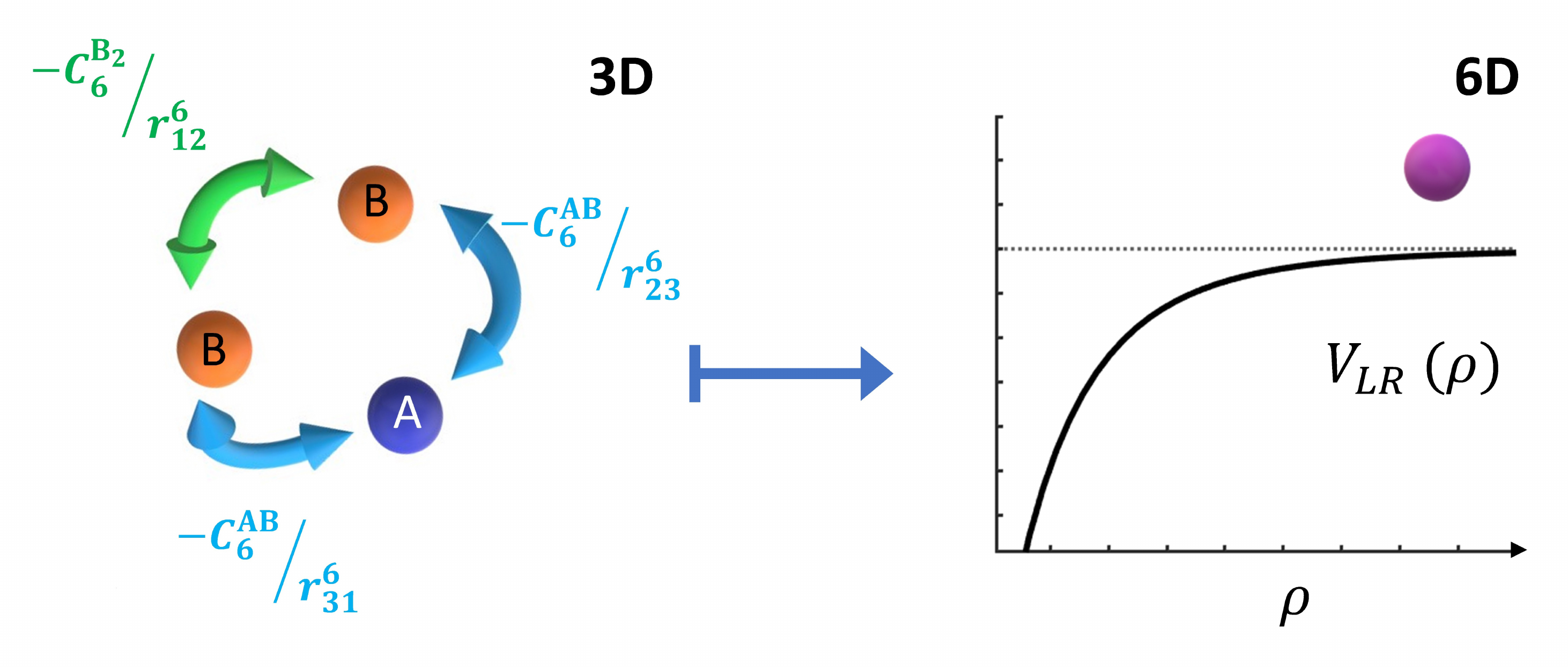}
			\caption{A schematic illustration of the long-range vdw interaction between three particles in 3D space and its counterpart, $V_{LR}(\rho)$, for a single particle in the 6D space.}	
			\label{fig:map}	
		\end{center}
	\end{figure*}
	
	\subsection{Potential}\label{subsec:pot}
	Throughout the present work we make use of the pairwise additive approximation which states that the total potential of a $N$-body system is the sum of all two-body interactions in the system. Thus, introducing the pairwise potentials $U(r_{ij})$, where $r_{ij} = |\vec{r}_j - \vec{r}_i|$, we write the interaction potential $V$ in \cref{eq:cartesianH} in the following form
	\begin{equation}\label{eq:add_pot}
		V(\vec{r}_1,\vec{r}_2,\vec{r}_3) = U(r_{12}) +  U(r_{23}) +  U(r_{31}) ~.
	\end{equation}
	
	It is known that the pairwise-additive descriptions of vdW interactions provide appropriate results for the calculation of crystal binding energies~\cite{Elrod1994} (showing deviations $\lesssim 10\%$), long-range coefficients of small molecules,~\cite{Reilly2015} and the spectroscopy of clusters,~\cite{MoazzenAhmadi2013} although in the latter case it is used only for its convenience. However, there are some scenarios in which a many-body interaction term is required, namely the calculation of long-range coefficients in large molecules~\cite{Reilly2015} and accurate spectroscopic constants of vdW complexes.~\cite{Elrod1994} On the contrary, scattering observables are accurately described at the ultracold limit without invoking many-body interaction terms in the underlying potential energy surface (see, e.g., Ref.~[\onlinecite{Makrides2015}]). Based on these examples and considering the nature of systems studied in this work, a pairwise approximation for the three-body potential $V(\vec{r}_1,\vec{r}_2,\vec{r}_3)$ is convenient.

	The relative distances $r_{ij}$ in Cartesian coordinate are related to the Jacobi vectors through the following equations  
	\begin{align}\label{eq:rel2Jac}
		r_{12} & = \left|\vec{\rho}_1\right| ~, \nonumber \\
		r_{23} & = \left|\vec{\rho}_2- \frac{m_1}{m_1+m_2}\vec{\rho}_1 \right| ~, \nonumber \\
		r_{31} & = \left|\vec{\rho}_2+ \frac{m_2}{m_1+m_2}\vec{\rho}_1 \right| ~.
	\end{align}
	Using these relations together with \cref{eq:rho6D} we can obtain the potential $V(\vec{\rho})$ in \cref{eq:6DH} from \cref{eq:add_pot}. It is important to emphasize that, due to the relations given by \cref{eq:hsCart}, potential is a function of the magnitude of the 6D position vector, $\rho$, and the corresponding hyperangles $\boldsymbol{\alpha} \equiv (\alpha_1,\alpha_2,\alpha_3,\alpha_4,\alpha_5)$, in other words, $V(\vec{\rho}) \equiv V(\rho,\boldsymbol{\alpha}) $. 
	
	\subsection{Grand angular momentum}\label{subsec:GAM}
	To finalize this section, let us briefly explain the notion of grand angular momentum in hyperspherical coordinates. Further below, we will use this discussion to develop a capture model in the hyperspherical coordinate system, and with it a classical threshold law which is the main purpose of this work. In classical mechanics, angular momentum in 6D space is a bivector defined by the exterior product (also known as wedge product) of 6D position and momentum vectors as
	\begin{equation}\label{eq:grandL}
		\mathrm{\Lambda} = \vec{\rho} \wedge \vec{P}, 
	\end{equation} 
	which is isomorphic to a $6\times6$ skew-symmetric matrix with elements
	\begin{equation}\label{eq:compL}
		\Lambda_{ij} = \rho_i P_j - \rho_j P_i, 
	\end{equation}
	for $i,j = 1,2,\dots,6$. This general definition applies in all higher-dimensional spaces, and for the 3D space it coincides with the familiar cross product. It is worth mentioning that, even though $\mathrm{\Lambda}$ is not equal to the ordinary total angular momentum of the three-body system in the 3D space, it contains the components of the angular momenta (associated with Jacobi vectors) among its elements. Following the original definition by Smith in Ref.~[\onlinecite{Smith1960}], $\mathrm{\Lambda}$ is often referred to as the grand angular momentum.  
	
	Note that in quantum mechanics, $\mathrm{\Lambda}$ is an operator. Its square, $\mathrm{\Lambda}^2$, is the quadratic Casimir operator of so(6) with eigenvalues $\lambda(\lambda+4)$, where $\lambda$ is a positive integer, and with hyperspherical harmonics as the corresponding eigenfunctions.~\cite{Dragt1965,Whitten1968,Avery2012}
	
	\section{Long-range potential in hyperspherical coordinates}\label{sec:1}
	Consider a three-body collision A + B + B, where A and B are neutral atoms in their ground electronic state. The interaction potential $U(r_{ij})$ consists of a short-range repulsive interaction (due to the overlap of closed-shell orbitals) and a long-range vdW tail 
	\begin{align}
		U(r_{ij}) \rightarrow -\frac{C_6}{r_{ij}^6} ~,
	\end{align}
	for $r_{ij}$ greater than the LeRoy radius. 
	
	In Ref.~[\onlinecite{Mirahmadi2021}], we have shown that the formation of vdW molecules through direct three-body recombination at collision energies lower than the dissociation energy of the product molecule ($E_c < D_e^\mathrm{AB}$) is insensitive to the short-range interaction and is dominated by the long-range tail of the potential. Note that we do not consider the contribution of higher order terms ($ 1/r^8, 1/r^{10} ,\dots$) in the long-range tail of the potential, since the effect of long-range interaction on formation of vdW complexes is mainly through the $ 1/r^6$ term.~\cite{Mirahmadi2021} Our goal is to find a general expression for the long-range interaction potential associated with the three-body collision A + B + B,  in a 6D space relevant for the classical trajectory method that we employ (see \cref{fig:map}). 
	
	Following the discussion in \cref{subsec:pot} the long-range potential in hyperspherical coordinates is obtained via the relation
	\begin{equation}\label{eq:vlr_hs}
		V_{LR}(\rho,\boldsymbol{\alpha}) = -\frac{C_6^\mathrm{B_2}}{r_{12}^6}-\frac{C_6^\mathrm{AB}}{r_{23}^6}-\frac{C_6^\mathrm{AB}}{r_{31}^6}~.
	\end{equation}
	To fix the coefficients in this equation, we made use of different atoms chosen from alkali metals, alkaline-earth metals, transition metals, pnictogens, chalcogens, halogens, and rare gases. These atoms together with the dispersion coefficients of the pairwise interactions between atoms A and B, $C_6^\mathrm{AB}$, and between two B atoms,  $C_6^\mathrm{B_2}$, are listed in \cref{tab1}. 
	
	\begin{table*}
		\caption{\label{tab1} Dispersion coefficients of pairwise potentials contributing in different three-body collisions A+B+B and their corresponding temperature-dependent three-body recombination rates calculated from \cref{eq:MBLANg} at $T = 4$~K, a relevant temperature for buffer gas cell experiments. All the dispersion coefficients are given in atomic units and $k_3(T)$ is given in units of cm$^6/s$ .}
		\begin{ruledtabular}
			\begin{tabular}{llccllcc}
				A-B-B	& $C^\mathrm{AB}_6$ & $C^\mathrm{B_2}_6$  &  $k_3(T= 4$K) [$\times 10^{-32}$] & A-B-B & $C^\mathrm{AB}_6$ & $C^\mathrm{B_2}_6$& $k_3(T= 4$K)  [$\times 10^{-32}$] \\
				\hline
				Li-He-He	& 22.51\footnotemark[1] & 1.35\footnotemark[1] & 2.99  & F-Ar-Ar & 33.44\footnotemark[5] & 64.3 & 1.64  \\
				Na-	& 23.77\footnotemark[1] &  & 2.67  & Cd-	& 173.6\footnotemark[6] & & 1.36 \\
				K-	& 39.47\footnotemark[2] &  & 2.69  & Hg-	& 129.9\footnotemark[6] & & 1.28 \\
				Be-	& 12.98\footnotemark[3] &  & 2.79  & Zn-  & 139.4\footnotemark[6] &  & 1.44 \\
				Ca-	& 36.59\footnotemark[2] &  & 2.68 & Li-Kr-Kr  & 255\footnotemark[3] & 129.6\footnotemark[3] & 1.99 \\
				Sr-	& 38.64\footnotemark[4] &  & 2.62 & Na- & 289\footnotemark[3] & & 1.53 \\
				N-	& 5.7\footnotemark[4] &  & 2.53 & K- & 444.2\footnotemark[2] &  & 1.40 \\
				O-	& 5.83\footnotemark[5] &  & 2.50 & Be- & 146\footnotemark[3] & & 1.82 \\
				As-	& 17.16\footnotemark[4] &  & 2.49 & Ca-  & 400\footnotemark[3] &  & 1.39 \\
				P-	& 14.69\footnotemark[4] &  & 2.54 & Sr- & 482.1\footnotemark[2] &  & 1.21 \\
				Ti-	& 27.61\footnotemark[1] & & 2.60 & O- & 64.77\footnotemark[5] & &  1.52 \\
				Cd-	& 24.93\footnotemark[6] & & 2.53 & F- & 47.53\footnotemark[5] &  & 1.44\\
				Hg-	& 18.92\footnotemark[6] & & 2.46 & Cd-	& 250.9\footnotemark[6] & & 1.12\\
				Zn-	& 20.1\footnotemark[6] & & 2.52 & Hg-	& 186.9\footnotemark[6] & & 1.02\\
				Li-Ne-Ne    & 46.4\footnotemark[3] & 6.38\footnotemark[3] & 2.09 & Zn-	& 201.2\footnotemark[6] & & 1.21 \\
				Na-	& 52.4\footnotemark[3] &  & 1.69 & Li-Xe-Xe & 404\footnotemark[3] & 285.9\footnotemark[3] &  1.94 \\
				K-	& 77.5\footnotemark[2] &  & 1.61 &  Na- & 460.8\footnotemark[2] &  & 1.48 \\
				Be-	& 27.7\footnotemark[3] &  & 1.92 &  K- & 698.1\footnotemark[2] & & 1.35 \\
				Ca-	& 74.8\footnotemark[3] &  & 1.60  &  Be- & 226\footnotemark[3] & & 1.77 \\
				Sr-	& 86.36\footnotemark[2]  &  & 1.49 & Ca- & 624\footnotemark[3] & & 1.33 \\
				O-	& 13.13\footnotemark[5] &  & 1.64 & Sr- & 750.3\footnotemark[2] & & 1.14 \\
				Cd-	&46.3\footnotemark[6] & & 1.40 &  O- & 103.4\footnotemark[5]  &  & 1.48  \\
				Hg-	& 34.99\footnotemark[6] & & 1.34 & F- & 79.4\footnotemark[5] & & 1.40 \\
				Zn-	& 37.29\footnotemark[6] & & 1.45 &  Cd-	& 385.7\footnotemark[6] & & 1.05\\
				Li-Ar-Ar & 174.1\footnotemark[2] & 64.3\footnotemark[3] & 2.24 & Hg- & 285.8\footnotemark[6] & & 0.94\\
				Na-	& 196.8\footnotemark[2]  &  & 1.75 & Zn-	& 308.9\footnotemark[6] & & 1.15\\	
				K-	& 317\footnotemark[3]&  & 1.64 & Be-Li-Li	& 467\footnotemark[3] & 1389\footnotemark[3] & 4.91\\
				Be-	& 101\footnotemark[3] &  & 2.05 & Be-Na-Na	& 522\footnotemark[2] & 1363\footnotemark[3] &  3.39 \\
				Ca-	& 276\footnotemark[3] &  & 1.62 & Be-Mg-Mg	& 364.9\footnotemark[2] & 629.6\footnotemark[2] &  3.07 \\
				Sr- & 327.1\footnotemark[2]  &  & 1.46 & Be-Be-Be	& 213\footnotemark[3] & 213 & 3.69 \\
				O-	& 43.88\footnotemark[5] & & 1.73 & He-He-He	& 1.35\footnotemark[1] & 1.35 & 2.74\\
			\end{tabular}
		\end{ruledtabular}
		\footnotetext[1]{Dispersion coefficient $C_6$ for alkali-Helium pairs is taken from Ref.~[\onlinecite{Kleinekathofer1996,Kleinekathofer1999}], for He-He from Ref.~[\onlinecite{Aziz1995}], and for Ti-He from Ref.~[\onlinecite{Mirahmadi2021}].}
		\footnotetext[2]{Dispersion  coefficient $C_6$ is taken from Ref.~[\onlinecite{JIANG2015158}].}
		\footnotetext[3]{Dispersion  coefficient $C_6$ is taken from Refs.~[\onlinecite{Maeder1979,Kumar1985,Tao2010}].}
		\footnotetext[4]{Dispersion   coefficient $C_6$ for Sr-B is taken from  Ref.~[\onlinecite{Yin2010}] and for pnictogens-He from Ref.~[\onlinecite{Partridge2001}].}
		\footnotetext[5]{Dispersion  coefficient $C_6$ for O-B is taken from Ref.~[\onlinecite{Aquilanti1988}] and for F-B from Ref.~[\onlinecite{Aquilanti1988a}].}
		\footnotetext[6]{Dispersion coefficient $C_6$ obtained through the London-Drude theory of dispersion interactions and Taken from Ref.~[\onlinecite{Koperski2002}].}
	\end{table*}
	
	\subsection{$\rho$-distribution}\label{subsec:HRdist}
	To find the radial dependence of the potential $V_{LR}(\rho,\boldsymbol{\alpha})$, referred to as $V_{LR}(\rho)$ in \cref{fig:map}, we set the right-hand-side of \cref{eq:vlr_hs} equal to a constant, $-Q$, and solve the equation,  
	\begin{equation}\label{eq:rdist}
		\frac{C_6^\mathrm{B_2}}{r_{12}^6(\rho,\boldsymbol{\alpha})}+\frac{C_6^\mathrm{AB}}{r_{23}^6(\rho,\boldsymbol{\alpha})}+\frac{C_6^\mathrm{AB}}{r_{31}^6(\rho,\boldsymbol{\alpha})}	= Q ~,
	\end{equation}
	for randomly sampled hyperangles. Different hyperangles $\alpha_j$ are generated by means of the probability density function associated with $d\Omega$ given in \cref{eq:hsvol} to generate random points uniformly distributed on the 6-sphere (in the geometrical sense). This procedure implies that the solution of \cref{eq:rdist} will be obtained as a distribution of $\rho$ values, $f(\rho)$, for each particular $Q$. Finally, we choose $\rho = \rho_m$ with the maximum likelihood in the $\rho$-distribution, as a single value solution of \cref{eq:rdist}. 
	We select $Q$ from interval $[0.001, 1000]$ K and solve the equation for $10^4$ randomly generated sets of hyperangles. 
	Note that, even though for each set of $\boldsymbol{\alpha}$, \cref{eq:rdist} will be transformed into a sixth-degree equation of the variable $\rho$, we have seen that (regardless the value of $Q$) four of the roots are complex and from the two remaining real roots only one is positive. 
	
	Interestingly enough, for all chosen values of $Q$ and dispersion coefficients, the probability density function (PDF) of the $\rho$-distribution, $F_f(\rho)$,  is described by the PDF of the generalized extreme value (GEV) distribution, i.e.,  
	\begin{equation}\label{eq:rhodist}
		F_f(\rho) = \frac{1}{\delta}\exp\left[-\left(1+\xi\frac{\rho-\beta}{\delta}\right)^{-\frac{1}{\xi}}\right]\left(1+\xi\frac{\rho-\beta}{\delta}\right)^{-1-\frac{1}{\xi}} 
	\end{equation}
	with $1+\xi\frac{\rho-\beta}{\delta}>0$. The GEV distribution is a family of continuous probability distributions developed within the extreme value theory.~\cite{Singh1998,Markose2011} It is parametrized with a shape parameter $\xi\neq0$, a location parameter $\beta$, and a scale parameter $\delta$. It is worth mentioning that here $\xi > 0$, which corresponds to the type II (also known as Fr\'echet distribution) of GEV distributions.~\cite{Markose2011} The parameters of GEV in \cref{eq:rhodist} are fitted yielding an uncertainty below 0.05$\%$. 
	
\begin{figure}
		\begin{center}
			\includegraphics[scale=0.4]{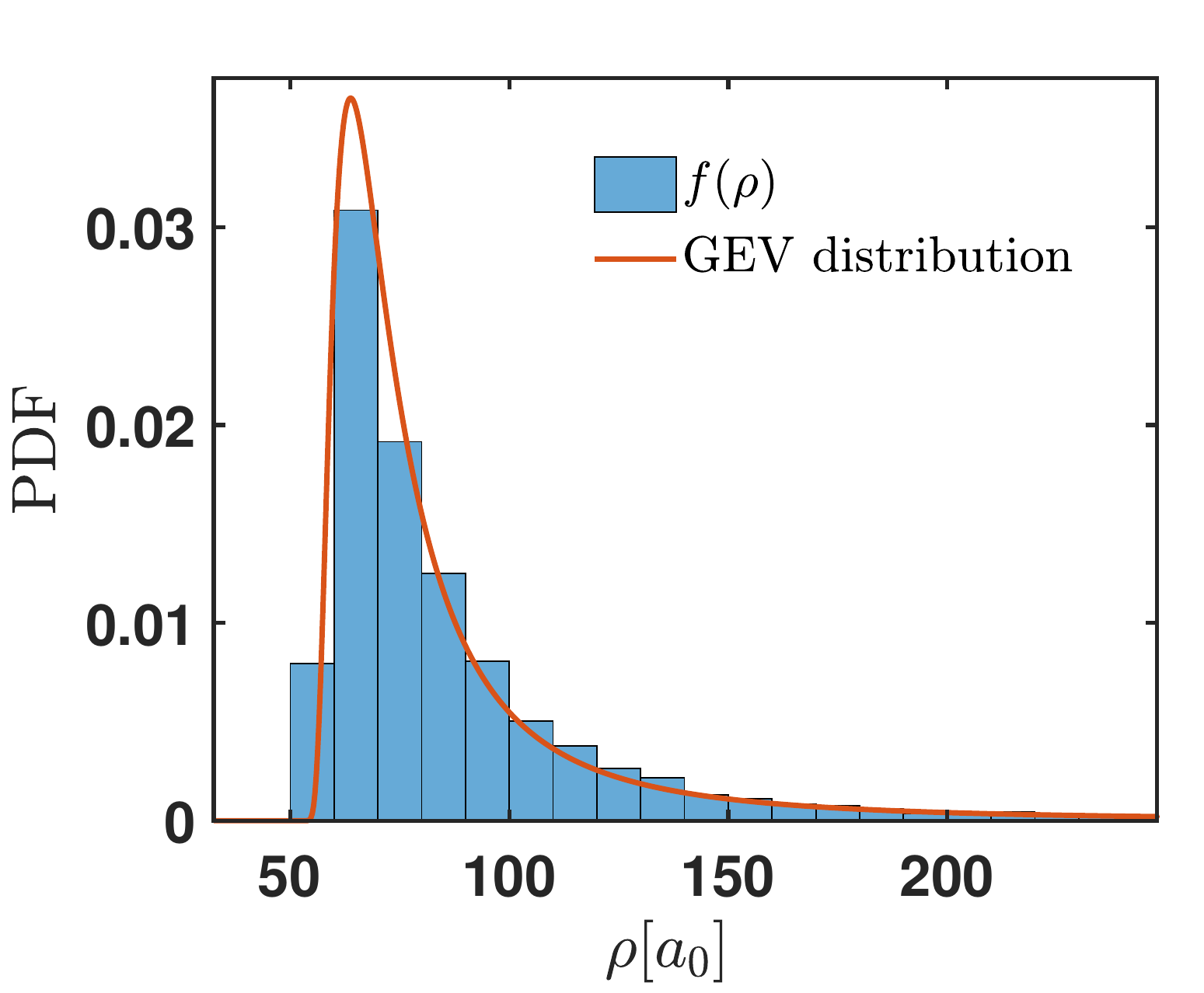}
			\caption{\label{fig:rhodist} The PDF of $\rho$-distribution $f(\rho)$ (blue histogram) and the fitted PDF of GEV distribution (red curve) for $Q = 1$ mK in the three-body collision As+He+He. $\rho$ is given in  units of the Bohr radius $a_0 \approx 5.29 \times 10^{-11} \mathrm{m}$. Parameters of GEV are obtained as $\xi \approx 0.67, \delta \approx 12.11$, and $\beta \approx 69.03$. }
		\end{center}
	\end{figure}

	As an illustration, the distribution $f(\rho)$ obtained by setting $Q = 1$ mK in \cref{eq:rdist} for the three-body collision As + He + He is shown in \cref{fig:rhodist}. As it can be seen in this figure, the probability density is positively-skewed. Therefore, due to this skewness, the maximum $\rho_m$ is achieved by using the mode of the GEV distribution,  
	\begin{equation}
		\rho_m = \beta + \frac{\delta}{\xi}\left(\left(1+\xi\right)^{-\xi}-1\right) ~,
	\end{equation}
	instead of its mean value. 		
	
	It is worth mentioning that we have observed the same (hyper)radial distribution $f(\rho)$ also in the case of pairwise interactions proportional to $1/r^4$. In other words, if we change an atom by an ion, the distribution of the hyperradius $\rho$ is of the same kind. However, the results of such a study are beyond the scope of this work and will be considered elsewhere. 
	
	\subsection{Effective dispersion coefficient}\label{subsec:c_eff}
\begin{figure}[h]
		\begin{center}
			\includegraphics[scale=0.4]{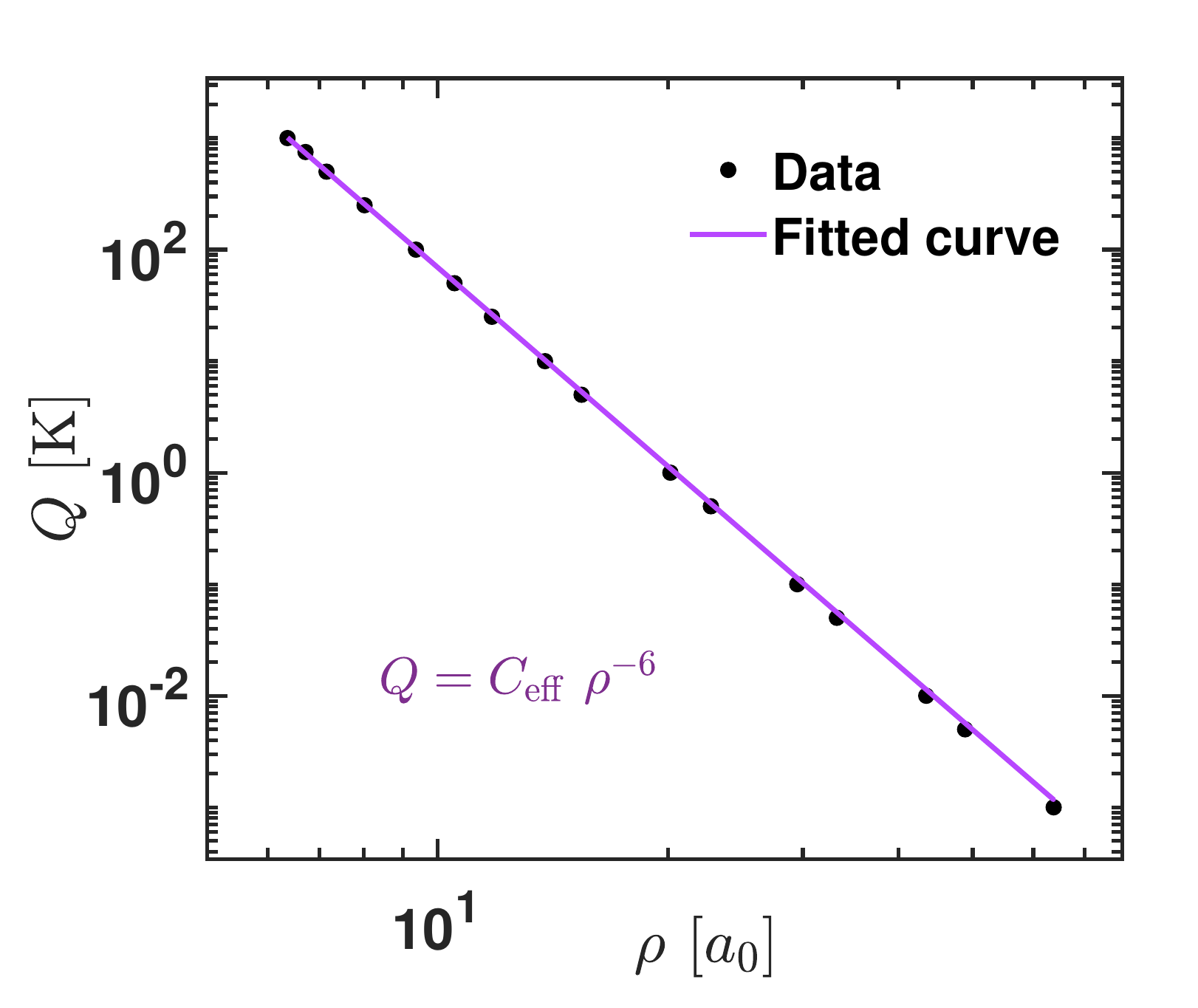}
			\caption{\label{fig:fittingQ} Power-law relation between different $Q$ and $\rho$ values (circles) for the three-body recombination process  As+He+He $\rightarrow$ AsHe+He  fitted with $Q = a \rho^{b}$ (purple line). The results are shown in the log-log scale.}
		\end{center}
	\end{figure}

	The general form of the long-range potential $V_{LR}(\rho)$ in the 6D space can be derived from the power-law relation between $Q$ and $\rho ~(=\rho_m)$. An example has been shown in \cref{fig:fittingQ}. It can be seen that, as expected, $Q$ is proportional to $1/\rho^6$ and the corresponding effective dispersion coefficient, $C_\mathrm{eff}$, is the slope of the fitted line in the log-log scale.  Therefore, $C_\mathrm{eff}$ can be obtained as a function of $C_6^\mathrm{AB}$ and $C_6^\mathrm{B_2}$ of the pairwise vdW interactions. Finally, utilizing \cref{eq:vlr_hs,eq:rdist},  we have 
	\begin{equation}\label{eq:vrad}
		V_{LR} (\rho)=	-\frac{C_\mathrm{eff}}{\rho^6} ~.
	\end{equation} 
	
\begin{figure}
		\begin{center}
			\includegraphics[scale=0.5]{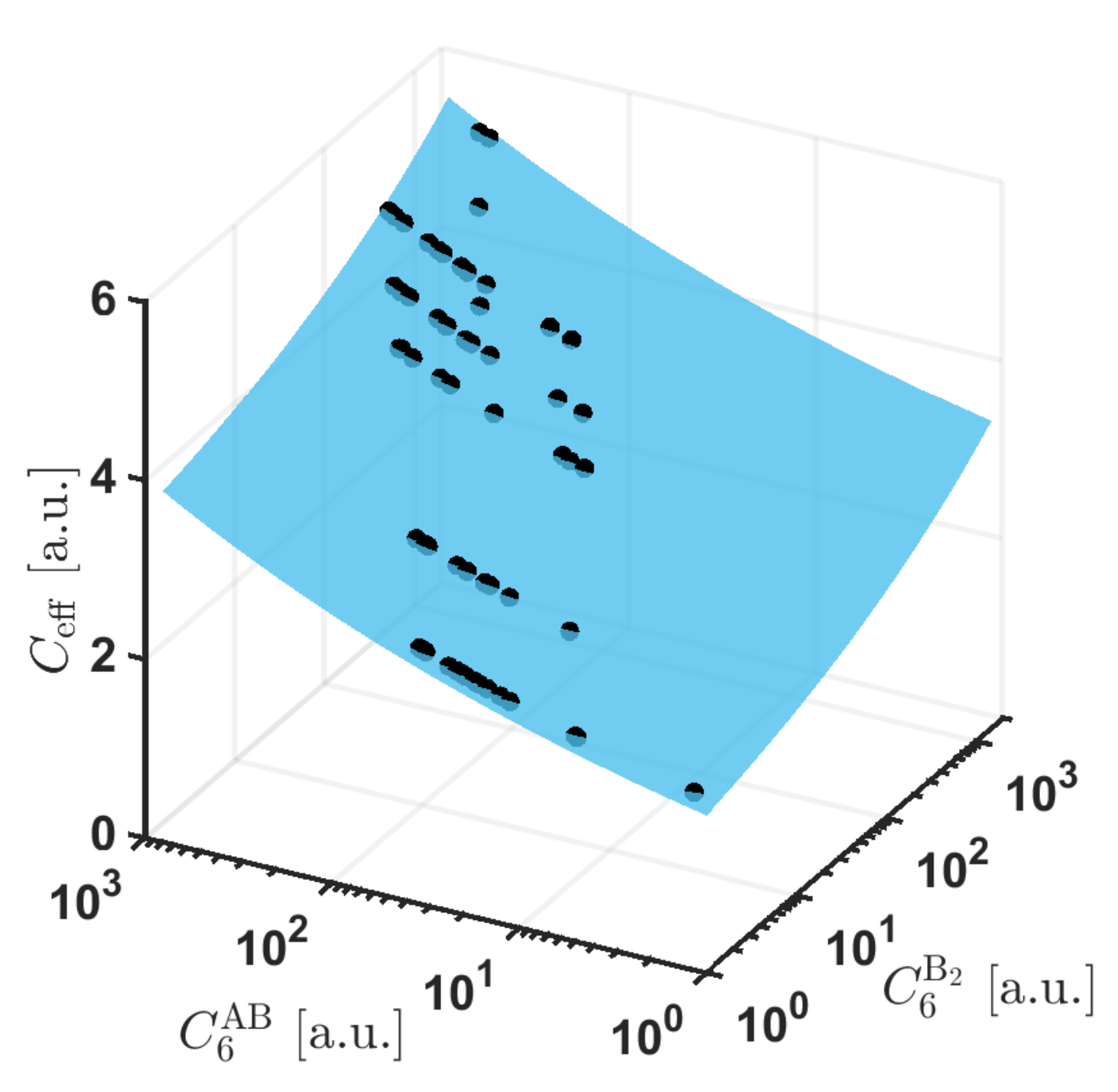}
			\caption{\label{fig:ceff} Calculated effective dispersion coefficient $C_\mathrm{eff}$ for different A+B+B collisions shown in \cref{tab1} (black symbols) together with $C_\mathrm{eff}$ calculated from \cref{eq:ceff} (blue surface). Note that the X and Y axis are in logarithmic scale. }
		\end{center}
	\end{figure}

	Through the same procedure for all A-B-B systems mentioned in \cref{tab1} and calculating the corresponding vdW coefficients $C_\mathrm{eff}$, we have found the general expression 
	\begin{equation}\label{eq:ceff}
		C_\mathrm{eff} = \mathrm{a}\left(C_6^\mathrm{AB}\right)^\mathrm{c} + \mathrm{d}\left(C_6^\mathrm{B_2}\right)^\mathrm{g},
	\end{equation}
	with parameters  $  \mathrm{a} = 0.56\pm 0.04$  , $\mathrm{c} = 0.189 \pm 0.009$ , $\mathrm{d} =  1.19\pm 0.04$ ,  and $ \mathrm{g} = 0.155\pm 0.003 $. \cref{eq:ceff} is applicable to any three-body system leading to the formation of vdW molecules. Note that $C_\mathrm{eff}$ is in atomic units and the value of parameters $\mathrm{a},\mathrm{b},\mathrm{c}$ and $\mathrm{d}$ should be modified for $C_\mathrm{eff}$ in other systems of units. \cref{fig:ceff} displays the surface plot of the coefficients obtained from  \cref{eq:ceff} with the mentioned parameters, plotted for different values of $C_6^\mathrm{AB}$ and $C_6^\mathrm{B_2}$.
	
	\section{Classical threshold law}\label{sec:2}
	A threshold law for the three-body recombination cross section and rate can be established based on the pioneering capture theory of Langevin.~\cite{Langevin1905} In the framework of this classical capture model, every trajectory associated with the collision energy ($E_c$) above the potential barrier leads, with unit probability, to a reaction event. To implement the same idea in 6D space, we first need to define the effective potential.

	In a three-body collision, after including the centrifugal energy, the effective long-range potential in hyperspherical coordinates reads as~\cite{Smith1960} (in atomic units) 
	\begin{equation}\label{eq:veff}
		V_\mathrm{eff}(\rho) = V_{LR}(\rho) + \frac{\mathrm{\Lambda}^2}{2\mu\rho^2}	~,
	\end{equation}
	with a maximum (the so-called centrifugal barrier) at $\rho_0 = \left(6\mu C_\mathrm{eff}/\mathrm{\Lambda}^2\right)^{1/4}$. Here, $\mathrm{\Lambda}^2 =  \left(\vec{\rho} \wedge \vec{P}\right)^2$ can be obtained from the components given by \cref{eq:compL} and after applying the (algebraic) Lagrange's identity~\cite{Gradshteyn2000} one finds
	\begin{align}\label{eq:length}
		\mathrm{\Lambda}^2 \equiv \sum_{1\leq i<j\leq6}\Lambda_{ij}^2 
		& = \left(\sum_{i=1}^6 \rho_i^2\right) \left(\sum_{i=1}^6 P_i^2\right)	-\left(\sum_{i=1}^6 \rho_i P_i  \right)^2 \nonumber \\ 
		& = \rho^2P^2 - \left(\vec{\rho}\cdot\vec{P}\right)^2 ~.
	\end{align} 
	Utilizing the relation between the impact parameter vector $\vec{b}$ and the initial position and momentum vectors,~\cite{Perez-Rios2014} we have $\vec{\rho}_0\wedge\vec{P}_0 = \vec{b}\wedge\vec{P}_0 $,  where we used the fact that $\vec{P}_0 \wedge \vec{P}_0 = 0$. 
	Finally, taking into account that $\vec{b}\perp\vec{P_0}$, one finds  
	\begin{align}\label{eq:Ebrel}
		\mathrm{\Lambda}^2 & = \left(\vec{b} \wedge \vec{P_0}\right)^2 \nonumber \\
		& = b^2 P_0^2  - \left(\vec{b}\cdot\vec{P_0}\right)^2 \nonumber \\
		& = 2\mu E_c  b^2, 
	\end{align}
	which establishes the intimate relation between the grand angular momentum and the magnitude of the impact parameter.

	Knowing that a reaction occurs only if  $E_c \sim V_\mathrm{eff}(\rho_0)$, we can find a threshold value for the impact parameter, $b_\mathrm{max}$, which is assigned to $E_c = V_\mathrm{eff}(\rho_0)$. Upon substituting for $\mathrm{\Lambda}^2$ obtained from \cref{eq:Ebrel} into $V_\mathrm{eff}(\rho_0)$, we derive the relation
	\begin{align}\label{eq:bmax}
		b_\mathrm{max} = \sqrt{\frac{3}{2}}  \left(\frac{2C_\mathrm{eff}}{E_c}\right)^{1/6} ~.
	\end{align}
	Inserting \cref{eq:bmax} into \cref{eq:sigma} we obtain the geometric cross section ($\mathcal{P}(E_c,b) = 1$ for $b\leq b_\mathrm{max}$) in the following form 
	\begin{align}\label{eq:sigmaLang}
		\sigma_\mathrm{rec}(E_c) & = \frac{8\pi^2}{3}\int_{0}^{b_\mathrm{max}(E_c)} b^4 d b \nonumber \\
		& = \frac{6\sqrt{3}\pi^2}{5\sqrt{2}} \left(2C_\mathrm{eff}\right)^{5/6} E_c^{-5/6} \nonumber \\
		& = \frac{6\sqrt{3}\pi^2}{5\sqrt{2}} \left[ 2\mathrm{a}\left(C_6^\mathrm{AB}\right)^\mathrm{c} + 2\mathrm{d}\left(C_6^\mathrm{B_2}\right)^\mathrm{g}\right]^{5/6} E_c^{-5/6} ~,
	\end{align}
	where in the last line we made use of \cref{eq:ceff}. Employing \cref{eq:k3}, the three-body recombination rate can be calculated as a function of collision energy and dispersion coefficients of the pairwise interactions as 
	\begin{equation}\label{eq:k3Lang}
		k_3(E_c) = \frac{6\sqrt{3}\pi^2}{5\sqrt{\mu}} \left[ 2\mathrm{a}\left(C_6^\mathrm{AB}\right)^\mathrm{c} + 2\mathrm{d}\left(C_6^\mathrm{B_2}\right)^\mathrm{g}\right]^{5/6} E_c^{-1/3} ~.
	\end{equation}
	Finally, the corresponding thermal average is obtained via integrating \cref{eq:k3Lang} over the appropriate three-body Maxwell-Boltzmann distribution of collision energies, yielding
	\begin{align}\label{eq:MBLANg}
		k_3(T) &=  \frac{1}{2(k_BT)^3} \int_{0}^{\infty}k_3(E_c) E_c^2 e^{-E_c/(k_BT)} dE_c ~, \nonumber \\
		& = \frac{4\pi^3 (k_BT)^{-1/3}}{3\Gamma(1/3)\sqrt{\mu}} \left[ 2\mathrm{a}\left(C_6^\mathrm{AB}\right)^\mathrm{c} + 2\mathrm{d}\left(C_6^\mathrm{B_2}\right)^\mathrm{g}\right]^{5/6} ~,
	\end{align}
	where $k_B$ is the Boltzmann constant and $\Gamma(x)$ is the gamma function of argument $x$.
	We should emphasize that these relations are best valid for collision energies smaller than the dissociation energy of the vdW molecules, which is typically below 100 K ($ \approx 10$ meV). Moreover, one should also verify the validity of the classical approach based on the number of involving partial waves, which will be discussed in what follows.

	\subsection{Estimated number of contributing three-body partial waves}\label{subsec:partial}
	From the perspective of a scattering problem of a single particle in a 6D space, each (grand) angular momentum quantum number $\lambda$ is a so-called partial wave. Following this fact, we introduce $\lambda$ as the partial wave associated with a three-body recombination in 3D space. 
	
	It is known that the reliability of the classical approach depends on the number of partial waves contributing to the scattering observables.~\cite{Perez-Rios2020,Perez-Rios2021} In other words, the large number of partial waves ($\approx 20$) contributing to the scattering washes out quantum effects such as resonances. The number of partial waves that impart a significant effect to the scattering problem can be estimated from the strength of interaction, i.e., the collision energy $E_c$.~\cite{Perez-Rios2020,Mirahmadi2021}
	
	Setting $\mathrm{\Lambda}^2$ equal to the eigenvalues of its quantum mechanical counterpart, $\lambda(\lambda+4)$, from $V_\mathrm{eff}(\rho_0) = E_c$ we can establish the following relation between the maximum three-body partial wave $\lambda_\mathrm{max}$ and a given collision energy 
	\begin{equation}\label{eq:numpar}
		\lambda_\mathrm{max} = \sqrt{6\mu} \left[ \mathrm{a}\left(C_6^\mathrm{AB}\right)^\mathrm{c} + \mathrm{d}\left(C_6^\mathrm{B_2}\right)^\mathrm{g}\right]^{1/6}\left(\frac{E_c}{2}\right)^{1/3}
	\end{equation} 
	where we made use of the fact that for $\lambda\gg4$, $\lambda(\lambda+4)\rightarrow \lambda^2$. \Cref{eq:numpar} provides a measure to check the validity of classical calculations for different A-B-B systems based on the collision energy, reduced mass, and pairwise dispersion coefficients. For a more detailed comparison between the quantum and classical results obtained by hyperspherical classical trajectory method in three-body recombination see Ref.~[\onlinecite{Perez-Rios2014}].		
	
	\subsection{Low-energy limit: $s$-wave collisions}\label{subsec:swave}
	In the final part of this section we derive a classical threshold law associated with the quantum $s$-wave scattering, i.e., $\lambda = 0$. In this case one may define the parameter $b_\mathrm{max}$ as the distance at which the collision energy is comparable to the strength of the interaction potential, i.e., $E_c = C_6^{\mathrm{B}_2} r_{12}^{-6} + C_6^{\mathrm{AB}} r_{23}^{-6} + C_6^{\mathrm{AB}} r_{31}^{-6} $ in 3D space or equivalently $E_c =  C_\mathrm{eff}\rho^{-6}$ in 6D space.     
	Therefore, the maximum impact parameter in the hyperspherical coordinate system reads as
	\begin{equation}\label{eq:bmaxHS}
		b_\mathrm{max} = \left(\frac{C_\mathrm{eff}}{E_c}\right)^{1/6} ~.
	\end{equation}
	By a similar argument as above, we obtain the cross section from \cref{eq:sigma}, which yields
	\begin{equation}\label{eq:sigmaHS}
		\sigma_\mathrm{rec}^\mathrm{sw}(E_c) = \frac{8\pi^2}{15} \left[ \mathrm{a}\left(C_6^\mathrm{AB}\right)^\mathrm{c} + \mathrm{d}\left(C_6^\mathrm{B_2}\right)^\mathrm{g}\right]^{5/6} E_c^{-5/6}.
	\end{equation}
	Consequently, the three-body recombination $k_3^\mathrm{sw}(E_c)$ and its thermal average $k_3^\mathrm{sw}(T)$ are given by 
	\begin{equation}\label{eq:k3HSN}
		k_3^\mathrm{sw}(E_c) = \frac{8\sqrt{2}\pi^2}{15 \sqrt{\mu}} \left[ \mathrm{a}\left(C_6^\mathrm{AB}\right)^\mathrm{c} + \mathrm{d}\left(C_6^\mathrm{B_2}\right)^\mathrm{g}\right]^{5/6} E_c^{-1/3}  
	\end{equation}
	and
	\begin{equation}\label{eq:MBHS}
		k_3^\mathrm{sw}(T) = \frac{16\sqrt{2}\pi^3\left(k_BT\right)^{-1/3}}{27\Gamma(1/3)\sqrt{3\mu}} \left[ \mathrm{a}\left(C_6^\mathrm{AB}\right)^\mathrm{c} + \mathrm{d}\left(C_6^\mathrm{B_2}\right)^\mathrm{g}\right]^{5/6},
	\end{equation}
	respectively. 
	
	\subsection{Results}\label{subsec:results}
	The results derived by performing the thermal average~\eqref{eq:MBLANg} for different A+B+B reactive collisions for $T=4$~K (relevant for buffer gas cells) are shown in \cref{tab1}. To calculate these values we used the mass of the most abundant isotopes of A and B atoms. Note that the calculated recombination rates account for both AB and B$_2$ products of the three-body process. However, based on the relative values of the dispersion coefficients, AB molecules are formed more often than B$_2$ ones, unless the dispersion coefficient for B$_2$ is larger than that of AB.
	It is important to notice that all calculated three-body recombination rates are of the same order of magnitude. One reason is the very close values of $C_\mathrm{eff}$ obtained for different systems, which almost neutralizes the effect of the three-body reduced mass $\mu$. 
	
	\cref{tab2} shows the three-body recombination rates $k_3(T)$ given by \cref{eq:MBLANg} for six different A+He+He collisions at $T = 4$~K, together with values of $k_3^\mathrm{num}(T)$ taken from Ref.~[\onlinecite{Mirahmadi2021}]. $k_3^\mathrm{num}(T)$ are the numerical values calculated via the classical trajectory method introduced in Ref.~[\onlinecite{Perez-Rios2014}]. The recombination rates $\tilde{k}_3(T)$ in this table are obtained from a capture model that only takes into account the pairwise interaction of the stronger long-range tail, i.e.,~\cite{Mirahmadi2021} 
	\begin{equation}\label{eq:k3old}
		\tilde{k}_3(T) = \frac{4\pi^3\left(k_BT\right)^{-1/3}}{\Gamma(1/3)\sqrt{\mu}}\left(2C_6^\mathrm{AB}\right)^{5/6} ~.
	\end{equation}
	
	It can be seen that the trend of the rates calculated with a capture model \`a la Langevin ($k_3$) is in reasonably good agreement with the trend of $ k_3^\mathrm{num}$ except for Li and Na. This is because the dissociation energy of LiHe and NaHe are below 2~K. Hence, the collision energies considered to calculate $k_3(T=4$~K) have reached the high-energy regime ($E_c>D_e$). Since this regime is sensitive to the short region of the potential, the capture model is not as accurate as in the other cases. Finally, we must highlight the considerable improvement (about one order of magnitude) in the threshold values $k_3$ obtained from \cref{eq:MBLANg}, over those values derived from the threshold law given by \cref{eq:k3old}. 
	
	\begin{table}
		\caption{\label{tab2} Temperature-dependent three-body recombination rates $k_3(T)$ from \cref{eq:MBLANg}, $\tilde{k}_3(T)$ from \cref{eq:k3old}, and $k_3^\mathrm{num}$ from Ref.~[\onlinecite{Mirahmadi2021}] calculated at $T = 4$~K. Recombination rates are given in units of [cm$^6/s\times10^{y}$] where $y$ is given in the parenthesis beside of each number.}
		\begin{ruledtabular}
			\begin{tabular}{cccc}
				A	& $k_3 $ & $\tilde{k}_3$  &  $k_3^\mathrm{num}$ \\
				\hline
				Li	& 2.99(-32)  & 2.03(-31) & 5.94(-33) \\
				Na	& 2.68(-32)  & 1.89(-31) & 9.30(-34)  \\
				N	& 2.53(-32)  & 5.99(-32) & 3.00(-33)  \\
				As	& 2.49(-32)  & 1.37(-31)& 2.94(-33)  \\
				P	& 2.54(-32)  & 1.25(-31) & 3.09(-33)  \\
				Ti	& 2.61(-32)  & 2.07(-31) & 3.09(-33)    \\
			\end{tabular}
		\end{ruledtabular}
	\end{table}
	
	\section{Conclusions and prospects}\label{sec:conclusion}
	
	After developing a clear picture of the (hyper)radial dependence of a three-body potential in a 6D space and studying more than 40 three-body systems relevant for vdW molecule formation, we have found how the long-range interaction of three-body systems depends on pairwise interactions between the colliding partners. Then, employing a classical trajectory method in hyperspherical coordinates, we have established a classical threshold law for the formation of vdW molecules through direct three-body recombination processes relevant for buffer gas cooling experiments. In addition, we have shown that at a given temperature, the three-body recombination rate is of the same order of magnitude independently of the atomic species under consideration, which corroborates our previous studies on the matter.
	
	The most valuable achievement of this work is to offer a simple expression which makes it possible to obtain the three-body recombination rate by only using the long-range dispersion coefficients and masses of the colliding atoms. This result helps to quickly estimate the role of three-body recombination in a given scenario, and with it, provides a new avenue for the calculation of three-body recombination rates avoiding costly computations. Finally, we hope that our findings help to make three-body collisions more approachable for the chemical physics community.

	\begin{acknowledgments}
		We would like to thank Miruna T. Cretu for valuable discussions and Gerard Meijer for his support and interest. 
	\end{acknowledgments}

	\bibliographystyle{apsrev4-2}
	\bibliography{trheshold_vdW_bib.bib}   
\end{document}